\documentclass[10pt,conference,epsfig]{IEEEtran}
\usepackage{graphicx}
\usepackage{subfigure,color}
\usepackage{amssymb}
\usepackage{cite,comment}
\usepackage{amsmath}
\usepackage{bm,comment}
\usepackage{algorithm}
\usepackage{algpseudocode}
\usepackage{geometry}
\makeatletter

\newcommand{\Rmnum}[1]{\expandafter\@slowromancap\romannumeral #1@}
\newcommand{\mv}[1]{\mbox{\boldmath{$ #1 $}}}

\makeatother
\newtheorem{fact}{Fact}

\newtheorem{definition}{Definition}

\begin{document}
\title{Cooperative Multi-Beam Routing for Multi-IRS Aided Massive MIMO}
\author{\IEEEauthorblockN{Weidong Mei\IEEEauthorrefmark{1}\IEEEauthorrefmark{2} and Rui Zhang\IEEEauthorrefmark{2}}
\IEEEauthorblockA{\IEEEauthorrefmark{1}NUS Graduate School, National University of Singapore}
\IEEEauthorblockA{\IEEEauthorrefmark{2}Department of Electrical and Computer Engineering, National University of Singapore}
Emails: wmei@u.nus.edu; elezhang@nus.edu.sg}
\markboth{IEEE WIRELESS COMMUNICATIONS}{}
\maketitle

\begin{abstract}
Intelligent reflecting surface (IRS) is envisioned to play a significant role in future wireless communication systems thanks to its powerful capability of enabling smart and reconfigurable radio environment. In this paper, we study the multi-IRS aided downlink communication in a massive multiple-input multiple-output (MIMO) system, where a multi-antenna BS simultaneously serves multiple remote single-antenna users with orthogonal beams reflected by multiple IRSs. By exploiting the line-of-sight (LoS) link between each pair of selected IRSs, a multi-hop cascaded LoS link can be established between the BS and each user via their cooperative beam routing. Under this setup, we optimize the selected IRSs and their beam routing path for each user, along with the BS/IRS active/passive beamforming such that the minimum received signal power among all users is maximized, subject to a new multi-beam routing path separation constraint for avoiding the inter-user/route interference. To tackle this problem, we first derive the optimal BS/IRS active/passive beamforming in closed-form for any given beam routes and show the beam routing optimization is NP-complete by recasting it as an equivalent graph-optimization problem. To solve this challenging problem, we then propose an efficient recursive algorithm to partially enumerate the feasible solutions, which effectively balances the performance-complexity trade-off by tuning its design parameter. Numerical results demonstrate that the proposed algorithm can achieve near-optimal performance with low enumeration complexity and also outperform other benchmark schemes.
\end{abstract}

\section{Introduction}
Wireless communications in the last several decades have witnessed remarkable progress in technology innovations such as channel coding, adaptive modulation, digital/analog beamforming etc. for significantly enhancing their performance by tens or even higher orders of magnitude. However, these techniques are only able to adapt to the random wireless channels but have limited control over them, thus leaving an ultimate barrier to achieving ultra-reliable and ultra-high-capacity wireless communication systems in the future. Recently, intelligent reflecting surface (IRS) has emerged as an appealing solution to tackle the above challenge. By dynamically tuning its massive reflecting elements, IRS is able to ``reconfigure'' the wireless channels and refine their realizations/distributions\cite{wu2019towards,wu2020intelligent}, rather than adapting to them only as for the traditional techniques. Thus, by efficiently integrating IRSs into future wireless networks, it is anticipated to bring a quantum-leap improvement in their capacity/performance over existing wireless systems.

\begin{figure}[!t]
\centering
\includegraphics[width=3in]{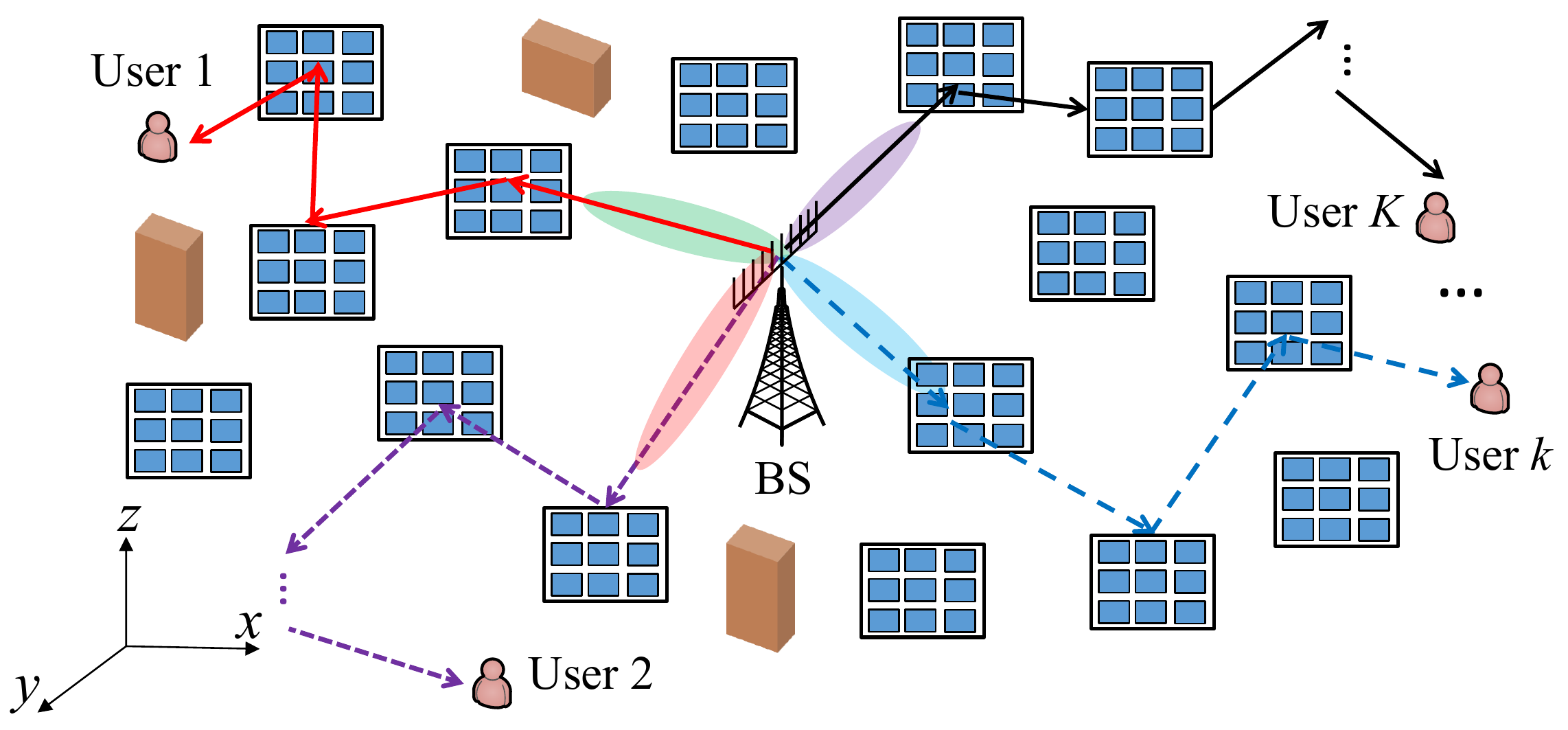}
\DeclareGraphicsExtensions.
\caption{A multi-IRS aided massive MIMO system with multi-beam routing.}\label{MultiBeam}
\vspace{-12pt}
\end{figure}
Due to the great potential of IRS, its performance has been extensively studied in the literature under different wireless system setups (see e.g.,\cite{wu2019towards,wu2020intelligent,hou2020reconfigurable,jiang2019over,pan2020intelligent,yu2020robust}). However, most of the existing works consider one or multiple distributed IRSs, which assist in the wireless communication between the base station (BS) and users with only one single signal reflection by the IRS per link. This simplified approach, however, generally results in suboptimal performance. This is because by carefully deploying the IRSs, strong line-of-sight (LoS) channel can be achieved for inter-IRS links, which can provide more pronounced cooperative passive beamforming (CPB) gains over the conventional single-IRS assisted system. Inspired by this, the authors in \cite{han2020cooperative} first proposed a double-IRS system, where a single-antenna BS serves a single-antenna user through a double-reflection link by two IRSs deployed near the BS and user, respectively. It was shown in \cite{han2020cooperative} that this system provides a CPB gain that increases {\it quartically} with the total number of IRS reflecting elements, compared to the {\it quadratic} growth of the passive beamforming gain in the conventional single-reflection link. The authors in \cite{you2020wireless} and \cite{zheng2020double} further extended \cite{han2020cooperative} to accommodate more practical Rician fading channel and multi-antenna/multi-user setup, respectively. Nonetheless, the general multi-IRS aided communication system with multi-hop (more than two hops) signal reflections, as shown in Fig.\,\ref{MultiBeam}, has not been addressed in the literature yet. Thanks to more available IRSs in the network, different end-to-end LoS routes can be achieved between the BS and multiple users at the same time by the multi-hop signal reflections of multiple sets of IRSs selected. This thus gives rise to a new {\it cooperative multi-beam routing} problem, where the selected IRSs and their beam-routing paths for different users are jointly optimized with the BS/IRS active/passive beamforming to maximize the minimum received signal power among all users. In our recent work\cite{mei2020cooperative}, by considering a single user in the system, we derived the optimal single-beam routing design.
 
In this paper, we study the multi-beam routing problem for the downlink communication in a massive multiple-input multiple-output (MIMO) system, where a multi-antenna BS serves multiple remote single-antenna users simultaneously over the same frequency band, aided by multiple distributed IRSs as shown in Fig.\,\ref{MultiBeam}. As compared to the single-beam routing design in \cite{mei2020cooperative}, a key challenge in the proposed cooperative multi-beam routing design is to avoid the inter-user/route interference due to undesired scattering by the IRSs that serve for different users/routes, especially when there exist LoS paths between them. This thus leads to a new and stringent multi-beam routing path separation constraint, where the IRSs selected for different routes should avoid having LoS channels between them. This constraint makes the multi-beam routing problem more challenging to solve, as the routing design needs to not only select the IRSs to maximize individual user's end-to-end cascaded LoS channel power, but also satisfy the path separation constraints for each pair of selected IRSs from different routes. To tackle this problem, we first derive the optimal BS/IRS active/passive beamforming in closed-form for any given beam-routing paths of the users, by exploiting the large BS antenna array induced orthogonal channels with IRSs and the inter-IRS LoS channels, respectively. Next, we show that the resultant multi-beam routing problem is NP-complete by recasting it into an equivalent neighbor-disjoint path optimization problem in graph theory. To deal with this challenging problem, an efficient recursive algorithm is proposed to partially enumerate the feasible solutions. By tuning its design parameter, the proposed algorithm is able to strike a flexible balance between performance and complexity. Numerical results show that the proposed algorithm can find the near-optimal multi-beam routing solutions with low enumeration complexity and also outperform other benchmark schemes.

{\it Notations:} $\lvert x \rvert$ and $\angle x$ denote the phase and amplitude of a complex number $x$, respectively. $({\mv a})_m$ denotes the $m$-th entry of a vector ${\mv a}$. $\lvert \Omega \rvert$ denotes the cardinality of a set $\Omega$. $\lfloor s \rfloor$ denotes the greatest integer not exceeding $s$.

\section{System Model and Beamforming Design}
\subsection{System Model}
As shown in Fig.\,\ref{MultiBeam}, we consider a massive MIMO downlink system, where $J$ distributed IRSs are deployed to assist in the communications from a multi-antenna BS to $K$ remote single-antenna users. Assume that the BS is equipped with $N \gg 1$ active antennas, while each IRS is equipped with $M$ passive reflecting elements. Due to the scattered obstacles, all BS-user direct links are assumed to be severely blocked. As such, the BS can only communicate with each user through a multi-reflection signal path which is formed by a set of IRSs associated with the user. To mitigate the scattered inter-user interference, the signal paths for all $K$ users should be sufficiently separated. Thus, each IRS is associated with at most one user. For convenience, we denote the sets of IRSs, IRS reflecting elements and users as ${\cal J}\triangleq \{1,2,\cdots,J\}$, ${\cal M}\triangleq \{1,2,\cdots,M\}$ and ${\cal K}\triangleq \{1,2,\cdots,K\}$, respectively. To maximize the reflected signal power by each IRS and ease the hardware implementation, we set the reflection amplitudes of each IRS to one. As such, the diagonal reflecting matrix of each IRS $j, j \in \cal J$ is given by ${\mv \Phi}_j={\rm diag}\{e^{j\theta_{j,1}},e^{j\theta_{j,2}},\cdots,e^{j\theta_{j,M}}\} \in {\mathbb C}^{M \times M}$. For convenience, we name the BS and user $k, k \in \cal K$ as nodes 0 and $J+k$ in the system, respectively. Accordingly, we define ${\mv H}_{0,j} \in {\mathbb C}^{M \times N}, j \in {\cal J}$ as the channel from the BS to IRS $j$, ${\mv S}_{i,j} \in {\mathbb C}^{M \times M}, i,j \in {\cal J}, i \ne j$ as that from IRS $i$ to IRS $j$, and ${\mv g}_{j,J+k}^{H} \in {\mathbb C}^{1 \times M}, j \in {\cal J}$ as that from IRS $j$ to user $k$.

As depicted in Fig.\,\ref{MultiBeam}, we assume that each IRS is equipped with a uniform rectangular array (URA) parallel to the $x$-$z$ plane, while the BS applies a uniform linear array (ULA). The antenna and element spacing at the BS and each IRS is assumed to be $d_A$ and $d_I$, respectively. The numbers of elements in each IRS's vertical and horizontal directions are assumed to be $M_1$ and $M_2$, respectively, with $M_1M_2=M$. Let $d_{i,j}, i \ne j$ denote the distance between nodes $i$ and $j$, for which some reference transmitting/reflecting elements of the BS/IRSs are selected without loss of generality. To ensure the far-field propagation between any two nodes, we assume that $d_{i,j} \ge d_0, \forall i \ne j$, where $d_0$ denotes the minimum distance to achieve this end. Then, by carefully deploying the $J$ IRSs, LoS propagation may be achieved between nodes $i$ and $j$ if $d_{i,j}$ is practically small (but larger than $d_0$). To describe the LoS availability between any two nodes $i$ (BS/IRS) and $j$ (IRS/user), we define a binary LoS condition indicator $l_{i,j} \in \{0,1\}$. In particular, $l_{i,j}=1$ indicates that the link between nodes $i$ and $j$ consists of an LoS link; otherwise, $l_{i,j}=0$. Furthermore, we set $l_{i,i}=0, \forall i$. Obviously, we have $l_{i,j}=l_{j,i}, \forall i,j$. Based on the LoS condition between any two nodes, a multi-hop LoS link can be established between the BS and each user $k, k \in \cal K$ by properly selecting a subset of associated IRSs. For example, if $l_{0,i}=l_{i,j}=l_{j,J+k}=1, i,j \in \cal J$, we can select IRSs $i$ and $j$ as the associated IRSs of user $k$, which successively reflect its intended signal from the BS toward its user. The IRSs that are not associated with any user in $\cal K$ are turned off to minimize the scattered interference in the network.

Next, we characterize the LoS channel between any two nodes in the system (if any), which is modeled as the product of array responses at two sides. Specifically, for the ULA at the BS, its array response is given by ${\mv a}_B(\vartheta) \in {\mathbb C}^{N \times 1}$ with $({\mv a}_B(\vartheta))_n=e^{-j2\pi(n-1)d_A\sin\vartheta/\lambda}$, where $\vartheta$ denotes its angle-of-departure (AoD) and $\lambda$ is the carrier wavelength. Whereas for the URA at each IRS, its array response is given by ${\mv a}_I(\vartheta^a,\vartheta^e) \in {\mathbb C}^{M \times 1}$ with $({\mv a}_I(\vartheta^a,\vartheta^e))_m=e^{-j2\pi d_I(\lfloor\! \frac{m-1}{M_1}\!\rfloor\sin\vartheta^e\cos\vartheta^a+(m-1-\lfloor\! \frac{m-1}{M_1}\!\rfloor M_1)\cos\vartheta^e)/\lambda}$, where $\vartheta^e$ and $\vartheta^a$ denote its elevation angle-of-arrival (AoA)/AoD and azimuth AoA/AoD, respectively. Furthermore, we define $\vartheta_{0,j}$ as the AoD from the BS to IRS $j$, $\varphi^a_{j,i}$/$\varphi^e_{j,i}$ as the azimuth/elevation AoA at IRS $j$ from node $i$, and $\vartheta^a_{i,j}$/$\vartheta^e_{i,j}$ as the azimuth/elevation AoD from IRS $i$ to node $j$. The above AoAs and AoDs can be estimated based on the geometric relationship of the BS, IRSs and users in the system\cite{han2020cooperative} or by integrating sensors to the IRSs. 

Based on the above information, we define ${\tilde{\mv h}}_{j,1}={\mv a}_B(\vartheta_{0,j})$ and ${\tilde{\mv h}}_{j,2}={\mv a}_I(\varphi^a_{j,0},\varphi^e_{j,0})$ for the LoS channel from the BS to IRS $j, j \in \cal J$, ${\tilde{\mv s}}_{i,j,1}={\mv a}_I(\vartheta^a_{i,j},\vartheta^e_{i,j})$ and ${\tilde{\mv s}}_{i,j,2}={\mv a}_I(\varphi^a_{j,i},\varphi^e_{j,i})$ for that from IRS $i$ to IRS $j, i,j \in \cal J$, and ${\tilde{\mv g}}_{j,J+k} = {\mv a}_I(\vartheta^a_{j,J+k},\vartheta^e_{j,J+k})$ for that from IRS $j$ to user $k, j \in {\cal J}, k \in \cal K$. Then, if $l_{0,j}=1$, the BS-IRS $j$ channel is expressed as
\begin{equation}\label{Ch1}
{\mv H}_{0,j} = \frac{\sqrt \beta}{d_{0,j}}e^{-\frac{j2\pi d_{0,j}}{\lambda}}{\tilde{\mv h}}_{j,2}{\tilde{\mv h}}^H_{j,1}, \;j \in {\cal J},
\end{equation}
where $\beta\, (<1)$ is the LoS path gain at the reference distance of 1 meter (m). Similarly, if $l_{i,j}=1, i,j \in \cal J$, the IRS $i$-IRS $j$ channel is given by
\begin{equation}\label{Ch2}
{\mv S}_{i,j} = \frac{\sqrt \beta}{d_{i,j}}e^{-\frac{j2\pi d_{i,j}}{\lambda}}{\tilde{\mv s}}_{i,j,2}{\tilde{\mv s}}^H_{i,j,1}, \;i \ne j, i, j \in {\cal J}.
\end{equation}
Finally, if $l_{j,J+k}=1$, the IRS $j$-user $k$ channel is expressed as
\begin{equation}\label{Ch3}
{\mv g}^H_{j,J+k} \!=\! \frac{\sqrt \beta}{d_{j,J+k}}e^{-\frac{j2\pi d_{{j,J+k}}}{\lambda}}{\tilde{\mv g}}^H_{j,J+k}, \;j \!\in\! {\cal J}, k \!\in\! {\cal K}.
\end{equation}
Based on (\ref{Ch1})-(\ref{Ch3}), we can characterize the multi-hop LoS channel between the BS and each user $k, k \in \cal K$, with any given reflection path and BS/IRS active/passive beamforming, as detailed next.

\subsection{Active/Passive Beamforming Design}
Let $\Omega^{(k)}=\{a^{(k)}_1,a^{(k)}_2,\cdots,a^{(k)}_{N_k}\}, k \in \cal K$ denote the reflection path from the BS to user $k$, where $N_k\, (\ge 1)$ and $a^{(k)}_n \in \cal J$ denote the number of associated IRSs for user $k$ and the index of the $n$-th associated IRS, with $n \in {\cal N}_k \triangleq \{1,2,\cdots,N_k\}$, respectively. For convenience, we define $a^{(k)}_0=0$ and $a^{(k)}_{N_k+1}=J+k, k \in \cal K$. Then, to ensure that each IRS in ${\cal N}_k$ only reflects user $k$'s information signal at most once, the following constraints should be met:
\begingroup
\allowdisplaybreaks
\begin{equation}\label{feasible1}
	a^{(k)}_n \in {\cal J}, \;a^{(k)}_n \ne a^{(k)}_{n'}, \forall n,n' \in {\cal N}_k, n \ne n', k \in {\cal K}. 
\end{equation}

Moreover, each constituent link of $\Omega^{(k)}$, along with the BS-IRS $a^{(k)}_1$ link and the IRS $a^{(k)}_{N_k}$-user $k$ link, should consist of an LoS link, i.e.,
\begin{equation}\label{feasible2}
	l_{a^{(k)}_n,a^{(k)}_{n+1}}=1, \forall n \in {\cal N}_k \cup \{0\}, k \in {\cal K}.
\end{equation}

Furthermore, to avoid the scattered inter-user interference, we consider that there is no direct LoS link\footnote{The methods and results in this paper are extendible to the more general path separation constraints, e.g., without $q$-hop LoS link between any two reflection paths, with $q \ge 1$.} between any two nodes belonging to different reflection paths (except the common node $0$ or the BS). Thus, we have
\begin{equation}\label{feasible3}
	l_{a^{(k)}_n,a^{(k')}_{n'}}=0, \;a^{(k)}_n \ne a^{(k')}_{n'}, \forall n,n' \ne 0, k,k' \in {\cal K}, k \ne k'.
\end{equation}

Thus, each $\Omega^{(k)}$ is a feasible route if and only if the constraints in (\ref{feasible1})-(\ref{feasible3}) are satisfied. Given $K$ feasible routes $\Omega^{(k)}, k \in \cal K$, we define $\mv w_k \in {\mathbb C}^{N \times 1}, k \in \cal K$ as the BS active beamforming design for user $k$. Then, the BS-user $k$ multi-reflection channel, denoted as $h_{0,J+k}(\Omega^{(k)})$, is expressed as
\begin{equation}\label{recvsig1}
{\mv g}^H_{a^{(k)}_{N_k},J+k}\!{\mv \Phi}_{a^{(k)}_{N_k}}\prod\limits_{n \in {\cal N}_k, n \ne N_k}\!\!\!\!\left(\!{\mv S}_{a^{(k)}_n,a^{(k)}_{n+1}}\!{\mv \Phi}_{a^{(k)}_n}\!\right)\!{\mv H}_{0,a^{(k)}_1}\mv w_k, k \in {\cal K},
\end{equation}
which depends on both the CPB design of the $N_k$ selected IRSs and the active beamforming design $\mv w_k$. By substituting (\ref{Ch1})-(\ref{Ch3}) into (\ref{recvsig1}), it can be shown that
\begin{equation}\label{recvsig2}
h_{0,J+k}(\Omega^{(k)})=e^{-j\phi_k}\kappa(\Omega^{(k)})\prod\limits_{n=1}^{N_k}A^{(k)}_n{(\tilde{\mv h}}^H_{a_1,1}\mv w_k), k \in {\cal K},
\end{equation}
where
\begin{equation}\label{recvsig3}
A^{(k)}_n = \begin{cases}
	{\tilde{\mv s}}^H_{a^{(k)}_1,a^{(k)}_2,1}{\mv \Phi}_{a^{(k)}_1}{\tilde{\mv h}}_{a_1^{(k)},2} &{\text{if}}\;\;n=1\\
	{\tilde{\mv s}}^H_{a^{(k)}_n,a^{(k)}_{n+1},1}{\mv \Phi}_{a^{(k)}_n}{\tilde{\mv s}}_{a^{(k)}_{n-1},a^{(k)}_n,2}&{\text{if}}\;\;2 \le n \le N_k-1\\
	\tilde{\mv g}^H_{a^{(k)}_{N_k},J+k}{\mv \Phi}_{a^{(k)}_{N_k}}{\tilde{\mv s}}_{a^{(k)}_{N_k-1},a^{(k)}_{N_k},2} &{\text{if}}\;\;n=N_k,
\end{cases}
\end{equation}
$\phi_k=\frac{2\pi}{\lambda} D(\Omega^{(k)})$ with $D(\Omega^{(k)})=\sum\nolimits_{n=0}^{N_k}d_{a^{(k)}_n,a^{(k)}_{n+1}}$ denoting the transmission distance from the BS to user $k$ under the route $\Omega^{(k)}$, and
\begin{equation}\label{pathgain}
\kappa(\Omega^{(k)})=\frac{(\sqrt\beta)^{N_k+1}}{\prod\limits_{n=0}^{N_k}d_{a^{(k)}_n,a^{(k)}_{n+1}}}
\end{equation}
denotes the cascaded end-to-end LoS path gain between the BS and user $k$ under $\Omega$, which turns out to be the product of the LoS path gains of all constituent links in $\Omega$.

It follows from (\ref{recvsig2}) and (\ref{recvsig3}) that to maximize each BS-user $k$ equivalent channel power, i.e., $\lvert h_{0,J+k}(\Omega^{(k)}) \rvert^2$, the magnitude of each $A^{(k)}_n$ in (\ref{recvsig2}) should be maximized. Accordingly, the phase shifts of each IRS $a^{(k)}_n, n \in {\cal N}_k$, should be set as
\begin{equation}\label{psall}
\theta_{a^{(k)}_n,m}\!=\!
\begin{cases}
	\angle(\tilde{\mv s}_{a^{(k)}_1,a^{(k)}_2,1})_m-\angle({\tilde{\mv h}}_{a_1^{(k)},2})_m &{\text{if}}\;n=1\\
	\angle(\tilde{\mv g}_{a^{(k)}_{N_k},J+k})_m-\angle({\tilde{\mv s}}_{a^{(k)}_{N_k-1},a^{(k)}_{N_k},2})_m &{\text{if}}\;n=N_k\\
	\angle(\tilde{\mv s}_{a^{(k)}_n,a^{(k)}_{n+1},1})_m-\angle({\tilde{\mv s}}_{a^{(k)}_{n-1},a^{(k)}_n,2})_m\!\!\! &{\text{otherwise}},
	\end{cases}
\end{equation}
for each $m \in \cal M$, leading to $\lvert A^{(k)}_n \rvert = M, \forall n \in {\cal N}_k, k \in \cal K$. 

Moreover, to maximize the magnitude of ${\tilde{\mv h}}^H_{a_1,1}\mv w_k$ in (\ref{recvsig2}), the BS active beamforming should apply the maximum-ratio transmission (MRT) based on ${\tilde{\mv h}}_{a_1,1}$. For ease of exposition, we assume that the BS equally allocates its total transmit power for the users in $\cal K$, and $\lvert w_k \rvert^2=1, k \in \cal K$. Then, its MRT is given by
\begin{equation}\label{bmall}
	{\mv w}_k = e^{j\phi_k}{\tilde{\mv h}}_{a^{(k)}_1,1}/{\lVert {\tilde{\mv h}}_{a^{(k)}_1,1} \rVert}, k \in {\cal K},
\end{equation}
and we have $\lvert {\tilde{\mv h}}^H_{a^{(k)}_1,1}\mv w_k \rvert = \sqrt{N}, \forall k \in \cal K$.

It is worth noting that if $N$ is sufficiently large, the BS active beamforming in (\ref{bmall}) ensures that the power of the information signal for each user $k, k \in \cal K$ overwhelms that of the inter-user interference in the BS-IRS $\footnotesize {a^{(k)}_1}$ link, i.e., the first link in $\Omega^{(k)}$. This is because with a large $N$, the BS antenna array has a practically high angular resolution. If all first-hop IRSs in the reflection paths $\Omega^{(k)}$'s, i.e., IRSs $a^{(k)}_1$'s, are sufficiently separated in the angular domain, the following asymptotically favorable propagation\cite{ngo2014aspects} can be achieved: 
\begin{align}
&\frac{1}{N^2}{\lvert\tilde{\mv h}}_{a^{(k)}_1,1}\rvert^4=\frac{1}{N}{\lvert\tilde{\mv h}}^H_{a^{(k)}_1,1}{\mv w}_k\rvert^2=1, k \in {\cal K}, \label{eq1}\\
&\frac{1}{N^2}{\lvert\tilde{\mv h}}^H_{a^{(k)}_1,1}{\tilde{\mv h}}_{a^{(k')}_1,1}\rvert^2\!=\!\frac{1}{N}{\lvert\tilde{\mv h}}^H_{a^{(k)}_1,1}{\mv w}_{k'}\rvert^2 \!\approx\! 0, k, k' \in {\cal K}, k \ne k'.\nonumber	
\end{align}
Thus, the inter-user interference is approximately nulled in the first link of each reflection path $\Omega^{(k)}$, by applying the MRT in (\ref{bmall}) with a large $N$. Furthermore, since the path separation constraints in (\ref{feasible3}) ensure that the scattered inter-user interference in the subsequent links of $\Omega^{(k)}$ is well mitigated, user $k$ is approximately free of inter-user interference, while achieving the maximum LoS channel power with the BS via the phase shifts in (\ref{psall}) and the BS's MRT in (\ref{bmall}).

By substituting (\ref{psall}) and (\ref{bmall}) into (\ref{recvsig2}), we obtain
\begin{equation}\label{eq2}
\lvert h_{0,J+k}(\Omega^{(k)}) \rvert^2=\frac{NM^{2N_k}\beta^{N_k+1}}{\prod\limits_{n=0}^{N_k}d^2_{a^{(k)}_n,a^{(k)}_{n+1}}}, k \in {\cal K}.
\end{equation}

It is observed from (\ref{eq2}) that for each BS-user $k$ equivalent channel, there exists a trade-off in maximizing the multiplicative CPB gain of $M^{2N_k}$ and maximizing the end-to-end path gain, i.e., $\kappa^2(\Omega^{(k)})$ in (\ref{pathgain}) (or minimizing the end-to-end path loss $\kappa^{-2}(\Omega^{(k)})$)\cite{mei2020cooperative}, as the former monotonically increases with $N_k$, while the latter generally decreases with $N_k$. Besides this trade-off, there also exists a fundamental trade-off in balancing all $\lvert h_{0,J+k}(\Omega^{(k)}) \rvert$'s for different users in $\cal K$. Specifically, due to the limited number of IRSs and LoS paths in the system as well as the path separation constraints in (\ref{feasible3}), maximizing the channel power for one user generally reduces the number of feasible routes for the other users. Particularly, if the number of users is large, some users may be denied access due to the lack of feasible routes. As such, the multi-beam routing should be properly designed to reconcile the above trade-offs, so as to ensure the optimum performance of all $K$ users.

\section{Problem Formulation}\label{pf}
In this paper, we aim to maximize the minimum signal-to-noise-plus-interference ratio (SINR) achievable by the $K$ users, by optimizing the reflection paths $\Omega^{(k)}, k \in \cal K$, subject to the feasibility constraints in (\ref{feasible1})-(\ref{feasible3}). Due to the well mitigated inter-user interference at each user's receiver, this is equivalent to maximizing the minimum BS-user LoS channel power, i.e., $\mathop {\min}\nolimits_{k \in \cal K} \lvert h_{0,J+k}(\Omega^{(k)})\rvert^2$. Thus, the optimization problem is formulated as
\begin{equation}\label{op1}
{\text{(P1)}} \mathop {\max}\limits_{\{\Omega^{(k)}\}_{k \in \cal K}}\;\mathop {\min}\limits_{k \in \cal K}\;\; \lvert h_{0,J+k}(\Omega^{(k)}) \rvert^2,\;\;\text{s.t.}\;\;{\text{(\ref{feasible1})-(\ref{feasible3})}}.
\end{equation}

However, (P1) is a combinatorial optimization problem due to its integer and coupled variables. Thus, it is very challenging to obtain the optimal solution to (P1) via standard optimization methods in general. To handle this challenging optimization problem, we next reformulate it as an equivalent graph-optimization problem. 

\section{Proposed Solution to (P1)}
In this section, we first reformulate (P1) as an equivalent problem in graph theory and thereby show that it is NP-complete. Then, a parametrized recursive algorithm is proposed to solve this problem.

\subsection{Problem Reformulation via Graph Theory}
Obviously, in (P1), it is equivalent to minimizing the maximum $\lvert h_{0,J+k}(\Omega^{(k)}) \rvert^{-2}$ among all $k \in \cal K$. Based on (\ref{eq1}), we have
\begin{equation}\label{eq3}
	\lvert h_{0,J+k}(\Omega^{(k)}) \rvert^{-2}=M^2\prod\limits_{n=0}^{N_k}\frac{d^2_{a^{(k)}_n,a^{(k)}_{n+1}}}{M^2\beta}, k \in {\cal K}.
\end{equation}

Next, we take the logarithm of (\ref{eq3}) and discard irrelevant constant terms. Then, (P1) becomes equivalent to
\begin{equation}\label{op2}
\mathop {\min}\limits_{\{\Omega^{(k)}\}_{k \in \cal K}}\mathop {\max}\limits_{k \in \cal K}\;\sum\limits_{n=0}^{N_k}\ln\frac{d_{a^{(k)}_n,a^{(k)}_{n+1}}}{M\sqrt\beta}, \\\quad\text{s.t.}\;\;{\text{(\ref{feasible1})-(\ref{feasible3})}}.
\end{equation}

To recast problem (\ref{op2}) as an equivalent problem in graph theory, we construct a directed weighted graph $G = (V,E)$. Specifically, the vertex set $V$ consists of all nodes in the system, i.e., $V=\{0,1,2,\cdots,J+K\}$. Furthermore, we consider that each of the $K$ beams can only be routed outwards from one IRS $i$ to a farther IRS $j$ from the BS with $d_{j,0} > d_{i,0}, i,j \in \cal J$, in order to reach its intended user as early as possible. Hence, the edge set $E$ is defined as
\begin{align}
E\!=&\{(0,j)| l_{0,j} \!=\! 1, j \!\in\! {\cal J}\}\!\cup\!\{(i,j)|l_{i,j} \!=\! 1, d_{j,0} \!>\! d_{i,0}, i,j \!\in\! {\cal J}\} \nonumber\\
&\cup \{(j,J\!+\!k)|\, l_{j,J+k} = 1, j \in {\cal J}, k \in \cal K\},\label{edgeSet}
\end{align}
i.e., there exists an edge from vertex $i$ to vertex $j$ if and only if an LoS path exists between them and $d_{j,0} > d_{i,0}$, except that vertex $j$ corresponds to a user, i.e., $j=J+k, k \in \cal K$. Note that (\ref{edgeSet}) ensures that there is no circle in $G$, i.e., $G$ is a direct acyclic graph (DAG). Moreover, the weight of each edge $(i,j)$ in $E$ is set as $W_{i,j}=\ln\frac{d_{i,j}}{M\sqrt\beta}$, which may be negative if $d_{i,j} < M\sqrt{\beta}$. Given the constructed graph $G$, any reflection path from the BS to user $k$ corresponds to a path from node 0 to node $J+k$ in $G$. Since $G$ is a DAG, any path in $G$ can automatically satisfy the constraints in (\ref{feasible1})-(\ref{feasible2}). To handle the more challenging constraint (\ref{feasible3}), we present the following definitions.
\begin{definition}
	Neighbor-disjoint paths refer to the paths in a graph which do not have any common or neighboring vertices except their starting points.
\end{definition}

According to Definition 1, the constraints in (\ref{feasible3}) can be satisfied if the $K$ paths from vertex $0$ to vertices $J+k, k \in \cal K$ are neighbor-disjoint. As such, problem (\ref{op2}) is equivalent to {\it finding $K$ neighbor-disjoint paths from vertex $0$ to vertices $J+k, k \in \cal K$ in $G$, respectively, such that the length of the longest path (i.e., the path with the maximum sum of edge weights) is minimized}. This problem is denoted as (P2). 

Note that neighbor-disjoint routing design has been previously studied in various multi-hop wireless networks, such as ad-hoc networks and wireless sensor networks, for the purpose of load balancing or interference mitigation\cite{teo2008interference,waharte2008probability}. However, most of these works only focused on discovering a set of neighbor-disjoint paths through different medium access control (MAC) layer protocols, but not from an optimal routing design perspective. A common routing design is by sequentially updating the routes for the $K$ users via the shortest path algorithm. After each update, the nodes in the optimized route and their neighbors in $G$ are removed before the next update, so as to satisfy (\ref{feasible3})\cite{teo2008interference}. However, this sequential update design generally yields suboptimal routes and even fails to return feasible routes, as the feasible set of routes for the current user critically depends on the routes for the previous users. In fact, it has been proved in \cite{waharte2008probability} that finding $K$ neighbor-disjoint routes in $G$ is NP-complete even in the case of $K=2$. As such, (P2) remains a challenging problem, which will be addressed next. 

\subsection{Proposed Solution to (P2)}\label{propsol}
The basic idea of the proposed solution is by first finding $Q\,(\ge 1)$ candidate shortest paths from node 0 to each node $J+k, k \in \cal K$. Given these candidate shortest paths, we construct a new {\it path graph}, based on which a recursive algorithm is performed to partially enumerate the feasible neighbor-disjoint paths and select the best one as the solution to (P2), as specified below.

{\it 1) Step 1: Find the candidate shortest paths.} First, for the nodes 0 and $J+k, k \in \cal K$, we invoke the Yen's algorithm\cite{west1996introduction} to find $Q$ candidate shortest paths between them. If the total number of paths between the two nodes is less than $Q$, we assume that there exist additional virtual paths between them with infinite sum of edge weights. For convenience, we denote by $p_k^{(q)}$ and $c_k^{(q)}, k \in {\cal K}, q \le Q$ the $q$-th shortest path between vertices 0 and $J+k$ and its sum of edge weights, respectively. Let $P=\{p_k^{(q)}, k \in {\cal K}, q \le Q\}$ be the set of all candidate shortest paths. The time complexity for this step is $O(KQ(J+K)(\lvert E \rvert+(J+K)\log (J+K)))$\cite{west1996introduction}.

{\it 2) Step 2: Construct the path graph.} Next, we construct a new undirected graph $G_p=(V_p,E_p)$, where each vertex in $V_p$ corresponds to one candidate shortest path obtained in Step 1 (thus termed as path graph), i.e., $V_p=\{v(p_k^{(q)})\,|\, k \in {\cal K}, q \le Q\}$. Hence, we have $\lvert V_p \rvert = KQ$. By this means, we can establish a one-to-one mapping between any path in $P$ and one vertex in $G_p$. Moreover, for any two vertices in $G_p$, there is an edge between them if and only if their corresponding paths in $G$ are neighbor-disjoint. Finally, we assign each vertex $v(p_k^{(q)})$ in $G_p$ with a weight, which is equal to the sum of edge weights of its corresponding path in $G$, i.e., $c_k^{(q)}$. It can be shown that the worst-case time complexity for this step is $O(Q^2K^2(J+K)^2)$.

To show the relationship between $G_p$ and $G$, we first introduce the following definitions.
\begin{definition}
	A $K$-partite graph refers to a graph whose vertices can be partitioned into $K$ disjoint sets, such that there is no edge between any two vertices within the same set.
\end{definition}
\begin{definition}
	A clique is a subset of vertices of an undirected graph, such that every two distinct vertices in the clique are adjacent. 
\end{definition}

Based on Definitions 2 and 3, we can verify the following facts, which specify the relationship between $G$ and $G_p$.
\begin{fact}
	$G_p$ is a $K$-partite graph, with the $k$-th disjoint set given by $V_{p,k}=\{v(p_k^{(q)}) \,|\,q \le Q\}, k \in \cal K$.
\end{fact}
\begin{fact}
	Among all paths in $P$, any $K$ neighbor-disjoint paths between vertices 0 and $J+k, k \in \cal K$, if they exist, correspond to a clique of size $K$ in $G_p$. 
\end{fact}

According to Facts 1 and 2, we aim to {\it find a clique of size $K$ in a $K$-partite graph $G_p$, whose maximum vertex weight is minimized}. This problem is denoted as (P3). The obtained clique corresponds to the best solution to (P2) among the paths in $P$. Thus, if $Q$ is set to be sufficiently large, such that all feasible paths from node 0 to each node $J+k, k \in \cal K$ are included in $P$, the proposed algorithm ensures to find an optimal solution to (P2), if (P3) is optimally solved. Accordingly, by tuning the value of $Q$, the proposed algorithm can flexibly balance between the performance and the complexity.

{\it 3) Step 3: Clique enumeration.} To find the optimal solution to (P3), we can enumerate all cliques of size $K$ in $G_p$ and then compare their respective maximum vertex weights. However, listing all cliques of size $K$ in a graph is also an NP-complete problem in general when $K > 2$\cite{west1996introduction}. Next, we propose a recursive algorithm to achieve this purpose by leveraging the $K$-partite property of $G_p$, thereby optimally solving (P3). 

Specifically, we show that each clique of size $K$ in $G_p$ can be recursively constructed based on the cliques of smaller sizes. Note that its $K$ vertices must be selected from the $K$ disjoint sets $V_{p,k}, k \in \cal K$, respectively. Without loss of optimality, we assume that its $k$-th vertex is selected from $V_{p,k}$. Accordingly, let $\Omega_r, r \le K$ denote the set of all cliques of size $r$ in $G_p$, with the $s$-th vertex of each clique selected from $V_{p,s}, s=1,2,\cdots,r$. Obviously, we have $\Omega_1=V_{p,1}$. Moreover, for each clique in $\Omega_r, r \le K-1$, if there exists a vertex in $V_{p,r+1}$ which is adjacent to all vertices in this clique, then a new clique in $\Omega_{r+1}$ can be constructed by appending the vertex to this clique. As such, based on the initial condition for $\Omega_1$ and the recursion for $\Omega_r, r \le K-1$, all feasible cliques of size $K$ can be enumerated, which incurs the worst-case complexity of ${\cal O}(Q^K)$. To further reduce the complexity, it is noted that when a clique of size $K\!-\!1$ is constructed, among all feasible vertices in $V_{p,K}$, we only need to append the vertex with the lowest weight to it. This is because the cliques resulted by appending other feasible vertices cannot yield a lower maximum vertex weight. Thus, the worst-case complexity of the recursive algorithm can be reduced to ${\cal O}(Q^{K-1})$. In fact, since the number of feasible vertices may significantly decrease when increasingly larger cliques are constructed (due to the more stringent adjacency constraint), the actual complexity of the proposed recursive algorithm is much lower than ${\cal O}(Q^{K-1})$, as will be shown in Section \ref{sim}. 

\section{Numerical Results}\label{sim}
\begin{figure}[!t]
\centering
\centering
\subfigure[]{\includegraphics[width=0.22\textwidth]{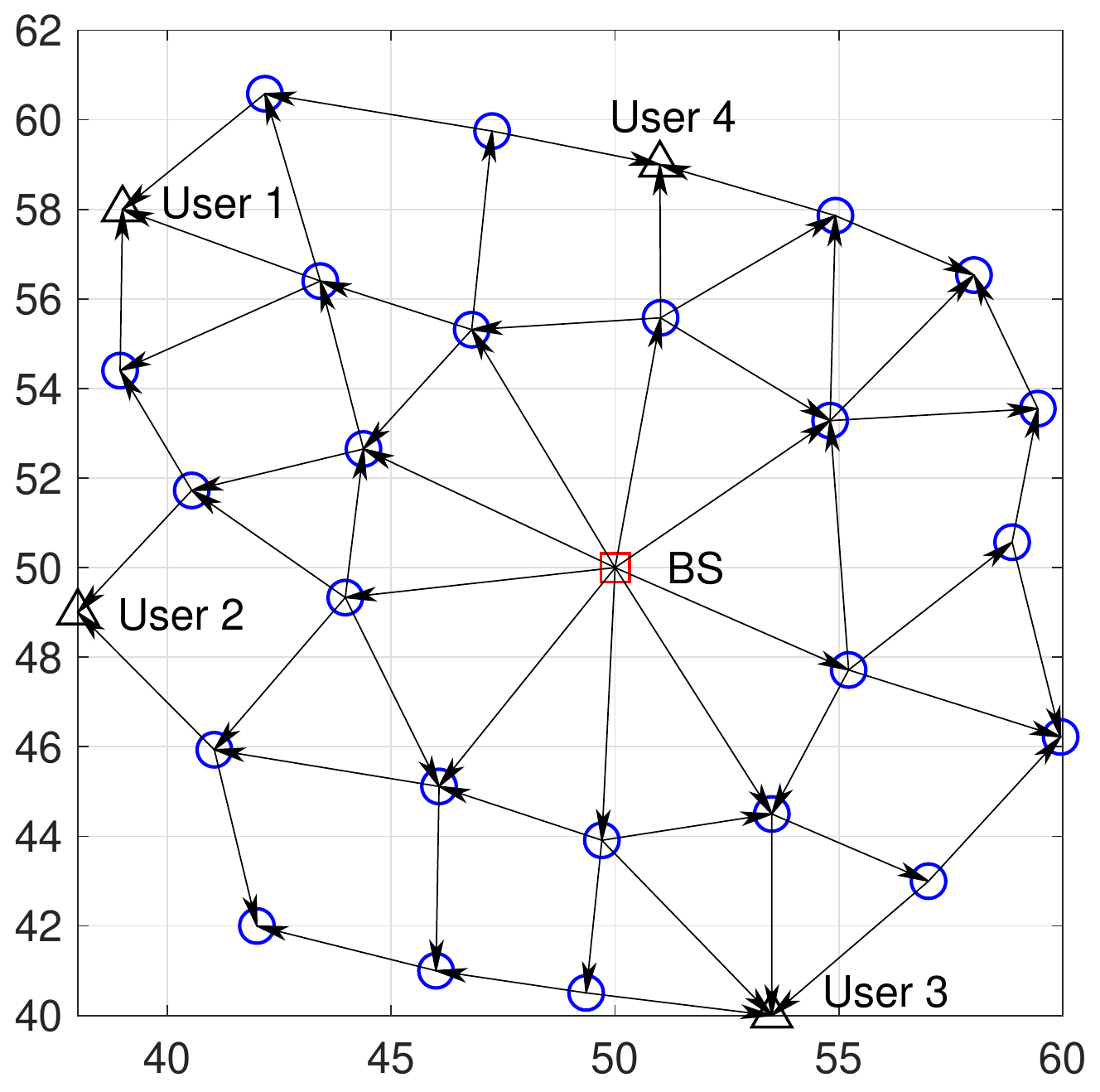}}
\subfigure[]{\includegraphics[width=0.222\textwidth]{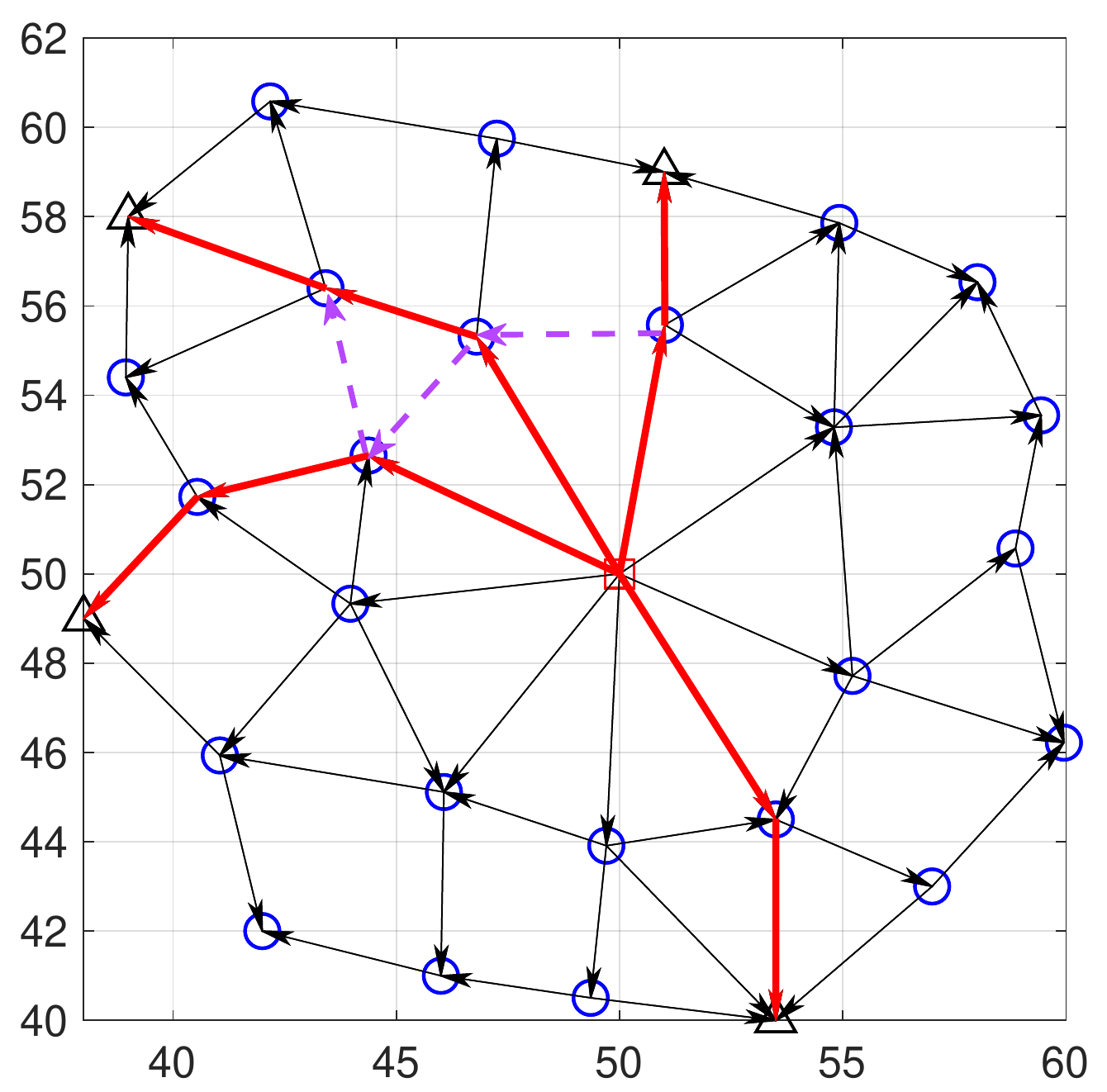}}
\caption{(a) Simulation setup (top view); (b) Optimal multi-beam routing without the path separation constraints in (\ref{feasible3}) under $M=400$.}\label{topology}
\vspace{-12pt}
\end{figure}
This section provides numerical results to validate our proposed multi-beam routing design. We focus on an indoor multi-IRS aided system (e.g., in a smart factory) operating at a carrier frequency of 5 GHz. Thus, the carrier wavelength is $\lambda=0.06$ m and the LoS path gain at the reference distance 1 m is $\beta=(\lambda/4\pi)^2=-46$ dB. Based on the LoS probability specified in \cite{3GPP38901}, we consider that there is an LoS link between two nodes $i$ and $j$, i.e., $l_{i,j}=1, i,j \in V$, if its occurrence probability is greater than 0.98, or $d_{i,j} \le$ 6.4 m. Moreover, we set the minimum distance for far-field propagation as $d_0=$ 3 m. Accordingly, the graph representation of the considered multi-IRS aided system, i.e., $G$, and the coordinates of all nodes (in (m)) are shown in Fig.\,\ref{topology}(a). The BS is equipped with $N=20$ antennas. It is verified via simulation that with the deployment of IRSs in Fig.\,\ref{topology}(a) and $N=20$, the asymptotically favorable propagation in (\ref{eq1}) can be achieved for all links between the BS (node 0) and the possible first-hop IRSs (the neighbors of node 0). In the proposed recursive algorithm, the number of candidate shortest paths for each node $J+k$ or user $k, k \in \cal K$ is set to $Q=20$. In Fig.\,\ref{topology}(b), by utilizing the Bellman-Ford algorithm\cite{west1996introduction} for the shortest path problem, we plot the optimal beam routing for each user without the path separation constraints in (\ref{feasible3}) under $M=400$. It is noted that there exist LoS links between the routes for users 1, 2 and 4, as highlighted in dashed lines, thus resulting in severe scattered inter-user interference. Thus, the proposed algorithm is needed to obtain a feasible multi-beam routing solution that meets (\ref{feasible3}).

\begin{figure}[!t]
\centering
\centering
\subfigure[]{\includegraphics[width=0.21\textwidth]{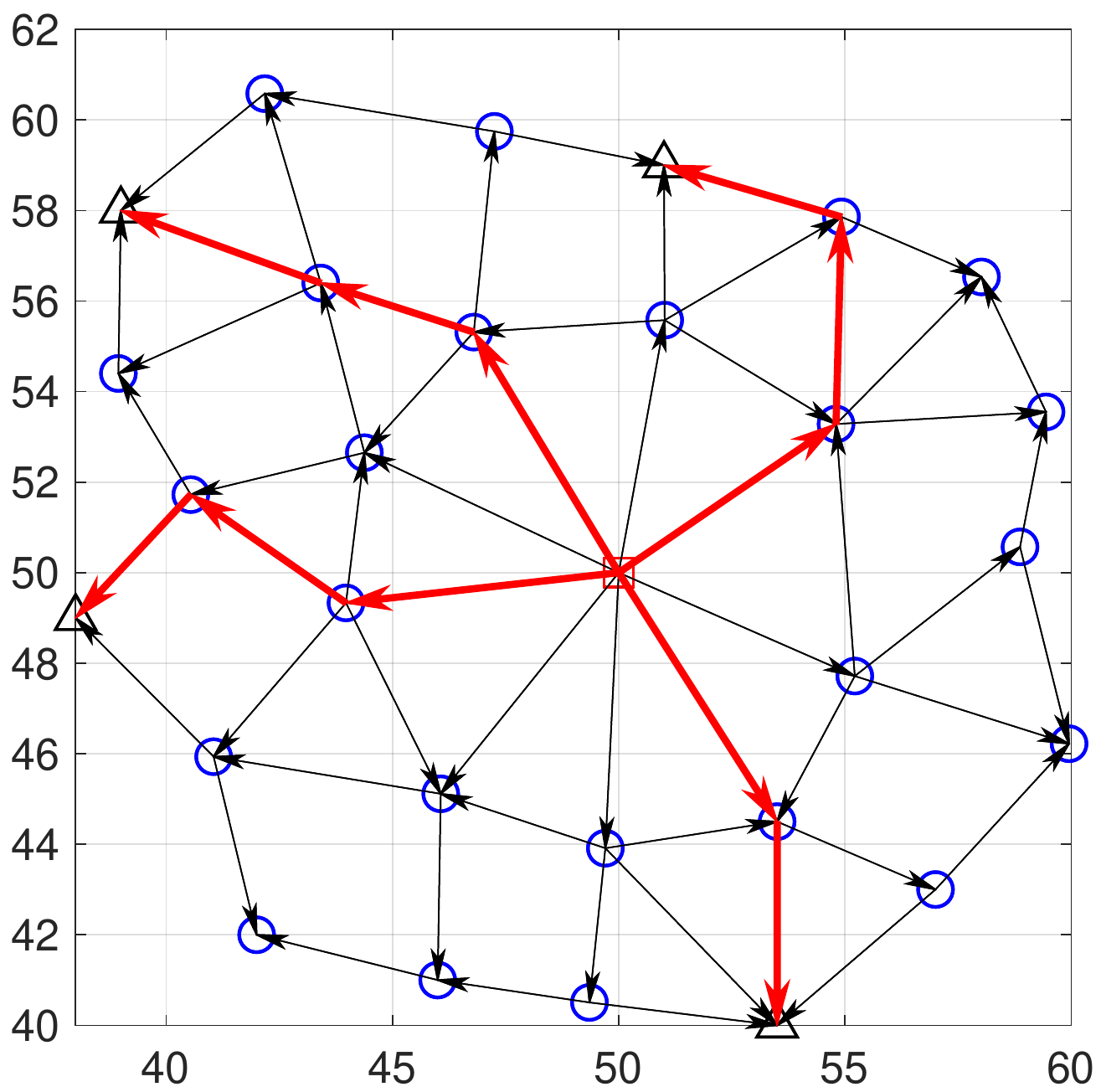}}
\subfigure[]{\includegraphics[width=0.208\textwidth]{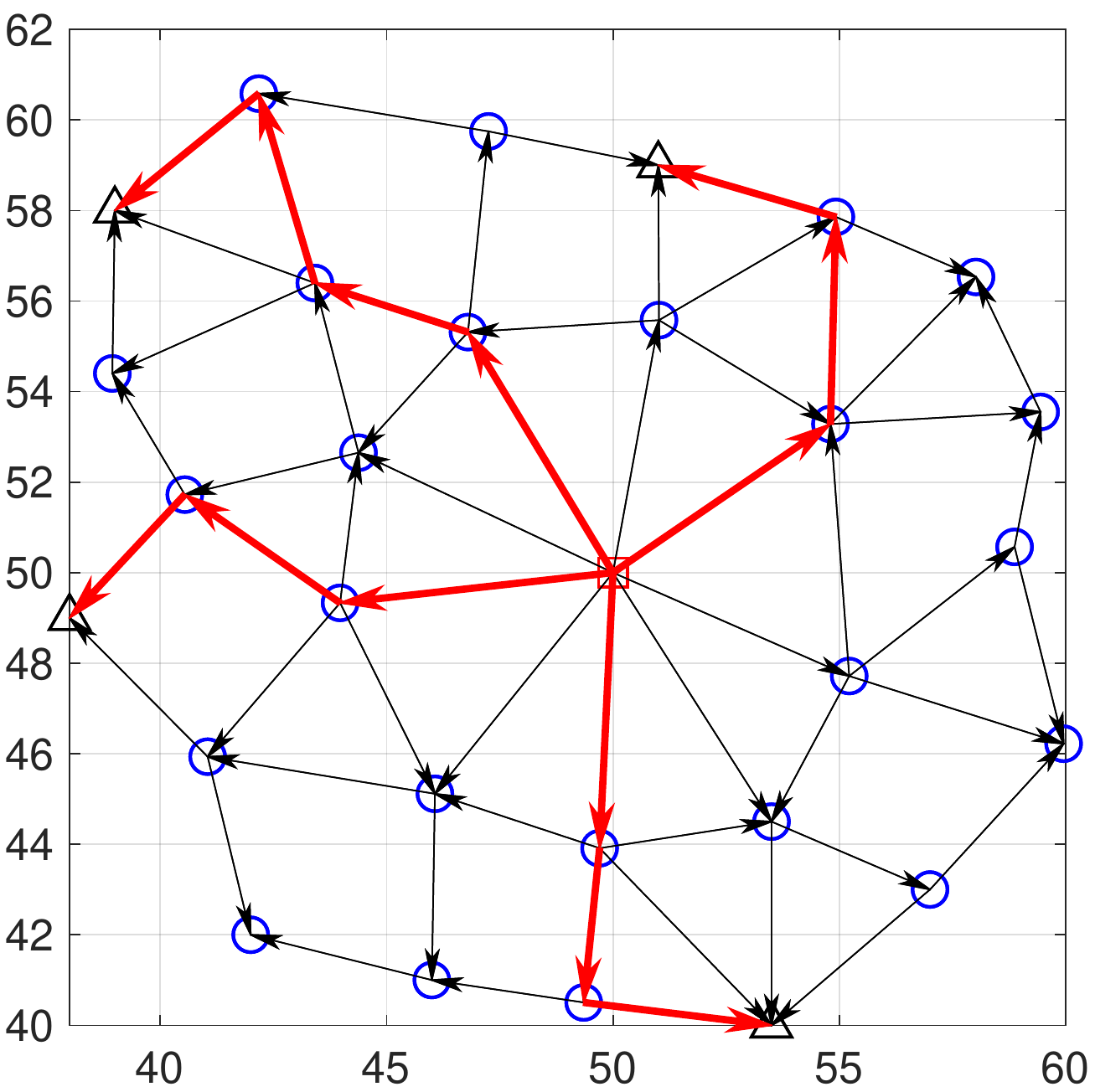}}
\vspace{-6pt}
\caption{Optimized routes with (a) $M=400$ and (b) $M=800$.}\label{optroute}
\vspace{-12pt}
\end{figure}
In Fig.\,\ref{optroute}, we plot the optimized multi-beam routing solution to (P1) under $M=400$ and $800$, respectively. By comparing Fig.\,\ref{optroute}(a) with Fig.\,\ref{topology}(b), it is observed that the routes for users 2 and 4 are changed due to the path separation constraints in (\ref{feasible3}). Accordingly, their SINR performance is sacrificed in order to balance the $K$ effective BS-user channel powers, i.e., $\lvert h_{0,J+k}(\Omega^{(k)})\rvert^2, k \in \cal K$, subject to (\ref{feasible3}). On the other hand, by comparing Fig.\,\ref{optroute}(a) with Fig.\,\ref{optroute}(b), it is observed that for users 1 and 3, their optimized routes under $M=800$ go through more IRSs than those under $M=400$. This indicates that as $M=400$, minimizing the end-to-end path loss is dominant over maximizing the CPB gain in maximizing their effective channel powers with the BS, and thus, the minimum BS-user channel power in (P1). While as $M$ increases, the CPB gain has a more significant effect in improving each $\lvert h_{0,J+k}(\Omega^{(k)})\rvert^2, k \in \cal K$ as compared to the end-to-end path loss. 

\begin{figure}[!t]
\centering
\centering
\subfigure[]{\includegraphics[width=0.24\textwidth]{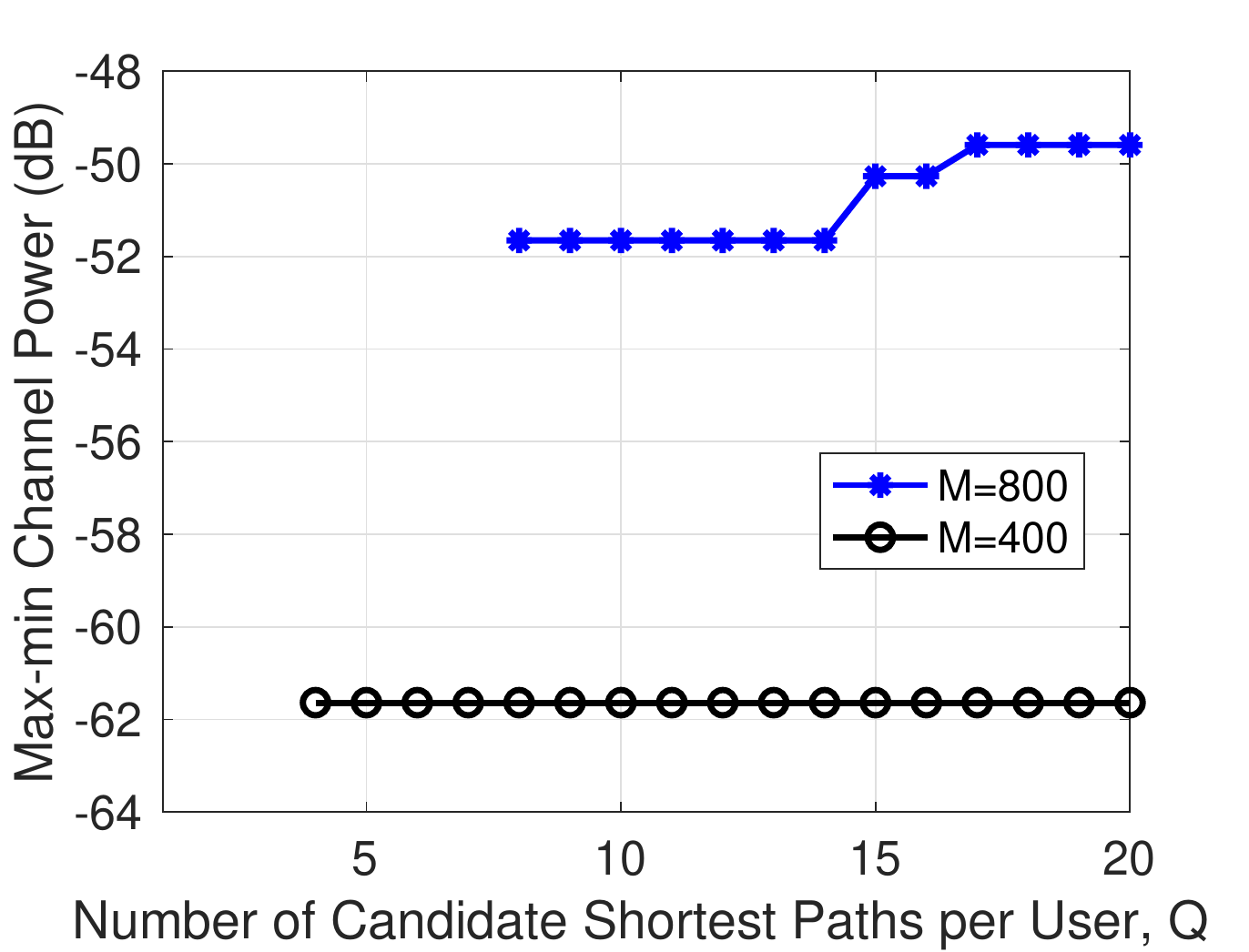}}
\subfigure[]{\includegraphics[width=0.24\textwidth]{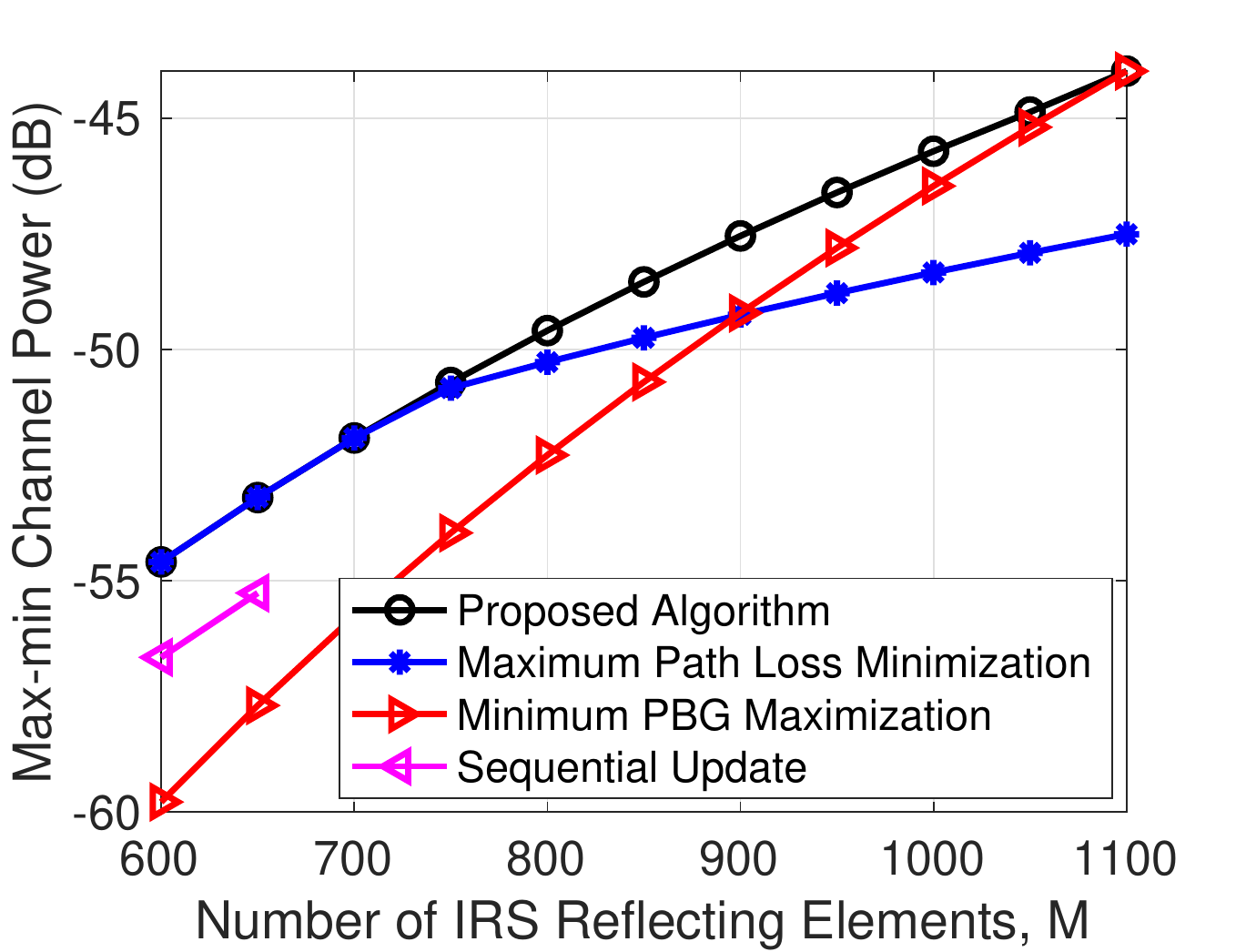}}
\vspace{-12pt}
\caption{Max-min channel power versus the number of (a) candidate shortest paths $Q$, and (b) IRS reflecting elements $M$.}\label{ChPwvs}
\vspace{-15pt}
\end{figure}
Fig.\,\ref{ChPwvs}(a) shows the maximized minimum (max-min) BS-user channel power among all users by the proposed recursive algorithm versus the number of candidate shortest paths per user $Q$, under $M=400$ and $M=800$. It is observed that the max-min BS-user channel power is monotonically non-decreasing with $Q$, since increasing $Q$ enlarges the size of the solution set of (P3). It can be verified that the performance of the proposed algorithm cannot be further improved by increasing $Q$ when $Q \ge 4$ and $Q \ge 17$ under $M=400$ and $M=800$, respectively. This implies that the optimal solution to (P2) (thus (P1)) is likely to be found by the proposed algorithm. It is also observed that a larger $Q$ is needed to find a feasible or a closer-to-optimal solution to (P3) as $M$ increases. The reason is that for any given $Q$, the $Q$ candidate shortest routes for each user generally go through more IRSs with increasing $M$ due to the more significant effect of CPB gain. As a result, the routes for different users in $\cal K$ are more likely to be close to each other and thus violate (\ref{feasible3}). Thus, $Q$ generally increases with $M$ to yield a feasible or better solution to (P3). Nonetheless, it is worth mentioning that even with a large $Q$ (e.g., $Q \ge 20$), the running time of the proposed clique enumeration is only around 0.06 seconds, which is very low for practical implementation.

Finally, Fig.\,\ref{ChPwvs}(b) shows the max-min BS-user channel power among all users by different approaches versus the number of IRS reflecting elements $M$. For performance comparison, we consider the following three benchmark schemes. The first scheme is the conventional sequential update scheme. As its performance critically depends on the order of the update for the users, we enumerate all possible $K!$ orders and show its best performance. The second benchmark minimizes the maximum path loss among all BS-user LoS links, while the third benchmark maximizes the minimum CPB gain among all BS-user LoS links. Their corresponding beam routes can be obtained by assuming $M=1$ and $M \rightarrow \infty$ in the proposed algorithm, respectively. It is observed from Fig.\,\ref{ChPwvs}(b) that the sequential update scheme fails to output feasible beam routes as $M > 650$. The second and third benchmarks are observed to achieve close performance to the proposed algorithm when $M \le 800$ and $M \ge 1000$, respectively, due to the different dominating effects of CPB gain and end-to-end path loss. However, when $800 < M < 1000$, these two schemes are observed to yield worse performance than the proposed algorithm, which strikes a better trade-off between maximizing the CPB gain and minimizing the end-to-end path loss. \vspace{-3pt}

\section{Conclusions}
This papers studies a new multi-beam routing problem for the multi-IRS aided massive MIMO system, where cascaded LoS links are established between the multi-antenna BS and multiple users by exploiting the successive signal reflections of selected IRSs. Under the stringent path separation constraints for avoiding the inter-user interference, the formulated problem is NP-complete and challenging to solve. To derive a high-quality suboptimal solution without incurring prohibitive complexity, we propose a parameterized recursive algorithm for this problem by leveraging graph theory. It is shown that the number of IRS reflecting elements has a great impact on the multi-beam routing solution as well as the achievable max-min user received signal power in the considered system.\vspace{-3pt}

\bibliography{IRScoop}
\bibliographystyle{IEEEtran}
\end{document}